\documentclass{kluwer}    

\input{psfig.tex}

\def\deg{$^{\circ}$}

\def\mmas{$\mu$as}

\def\teff{${\rm T}_{\rm eff}$}
\def\logg{$\log g$}

\def\fe{[Fe/H]}
\def\aabun{[$\alpha$/Fe]}
\def\vmicro{$V_{\rm micro}$}
\def\vrot{V$_{\rm rot}$}

\def\parallax{$\pi$}

\def\extinct{{\bf A}($\lambda)$}

\def\vv{{\bf v}}

\def\rv{{\bf r}}

\begin{document}
\begin{article}

\begin{opening}         

\title{Determination of stellar parameters with GAIA}
\author{C.A.L.\ \surname{Bailer-Jones}}  
\runningauthor{C.A.L.\ Bailer-Jones}
\runningtitle{Determination of stellar parameters with GAIA}
\institute{Max-Planck-Institut f\"ur Astronomie,
K\"onigstuhl 17, D-69117 Heidelberg, Germany. calj@mpia-hd.mpg.de
}
\begin{abstract}
The GAIA Galactic survey satellite will obtain photometry in 15
filters of over $10^9$ stars in our Galaxy across a very
wide range of stellar types. No other planned survey will
provide so much photometric information on so many stars. I examine
the problem of how to determine fundamental physical parameters
(\teff, \logg, \fe\ etc.) from these data. Given the size,
multidimensionality and diversity of this dataset, this is a
challenging task beyond any encountered so far in large-scale stellar
parametrization. I describe the problems faced (initial object
identification, interstellar extinction, multiplicity, missing data
etc.) and present a framework in which they can can be addressed. A
probabilistic approach is advocated on the grounds that it can take
advantage of additional information (e.g.\ priors and data
uncertainties) in a consistent and useful manner, as well as give
meaningful results in the presence of poor or degenerate
data. Furthermore, I suggest an approach to parametrization which can
use the other information GAIA will acquire, in particular
the parallax, which has not previously been available for large-scale
multidimensional parametrization.  Several of the problems identified
and ideas suggested will be relevant to other large surveys, such as
SDSS, DIVA, FAME, VISTA and LSST, as well as stellar parametrization
in a virtual observatory.
\end{abstract}

\end{opening}           

\vspace*{0.5em}
\begin{center}
\small{\em To appear in the proceedings of the GAIA conference, Vilnius July 2001\\
Ap\&SS, c.\ April 2002.}
\end{center}
\vspace*{-1.5em}

\section{Introduction}  

GAIA is an ESA-funded astrometric and photometric satellite for launch
in 2010--2012. It is similar in essence to Hipparcos, but extending by
several of orders of magnitudes the astrometric accuracy, magnitude
limit and number of objects. The goal is to measure positions to
10\mmas\ at V=15 and 160\mmas\ at V=20. It will observe the whole sky
down to V$\simeq$20 about 100 times over the course of four years,
observing an estimated $10^9$ stars, plus numerous galaxies, quasars
and asteroids. (For comparison, Hipparcos measured $10^5$ stars in
just one filter down to V=12 with an median astrometric accuracy of
1000\mmas\ for V<10.)  The astrometry will provide accurate positions,
distances and proper motions for many of these stars.  The primary
goal of GAIA is to explore the composition, formation and evolution of
the Galaxy by studying the dynamics and intrinsic properties of a wide
range of stellar types across the whole Galaxy.  GAIA will observe all
objects in 15 medium and broad band filters (FWHM between 10 and
200nm) across the wavelength range 250--950nm at a spatial resolution
of at least $0.5''$, although the optimization of this system
continues. High resolution (0.075nm) spectra of the brighter objects
will also be obtained (with a slitless spectrograph) in the vicinity
of Ca{\small II} triplet at 850--875nm, primarily to determine radial
velocities to a few km/s accuracy, thus complementing the astrometry
to give a 6D phase space map of all objects down to V$\sim$17. For
more details on the mission, see~\cite{esa_2000,perryman_etal_2001} or
Perryman's article in this volume.

In this article I examine the problem of determining fundamental
stellar parameters, e.g.\ \teff, R (radius), \fe\ and \aabun, with GAIA. As the
mission is still some ten years in the future, it would be premature
to present complete solutions to this problem. Rather, my emphasis is
on outlining the considerable challenges that this task
presents. The solution will have to rely extensively on automated
methods, which must be more robust and sophisticated than those used
to date for stellar classification and parametrization. These need to
deal with the complex distribution of objects in a multidimensional
colour or stellar parameter space: simply taking colour cuts to
produce two-dimensional colour-magnitude diagrams is an inefficient
(and often ineffective) use of multi-colour data.  After looking at
the scientific aspects of stellar parametrization and assessing what
is possible with GAIA (section~\ref{scireq}), I discuss the challenges
posed by the nature of this survey plus the practical
requirements which these place on parametrization methods
(section~\ref{pracreq}).  In section~\ref{models} I briefly mention
the capabilities and restrictions of some of the methods which have
been used so far on simpler stellar parametrization problems. Finally,
I present a framework approach for deriving stellar parameters with GAIA
(section~\ref{solution}).

\section{Scientific goals and capabilities}
\label{scireq}

The most fundamental properties of a star are its mass, age and
chemical composition. Unfortunately, age is not directly observable
and masses can only be determined directly (i.e.\ dynamically) in
select binary systems. Nonetheless, important parameters, in
particular the effective temperature, \teff, surface gravity, \logg,
and iron-peak metallicity, \fe, can be obtained from the stellar
spectral energy distribution (SED) given sufficient spectral
resolution. Combined with the parallax and interstellar extinction,
the luminosity, radius and mass can be determined. Also determinable
from the SED are the abundance of the alpha-process elements, \aabun,
CNO abundance anomalies, the microturbulence velocity, \vmicro,
rotational velocity, \vrot\ and activity.

\begin{table}[t]
\begin{tabular}{llll}
\hline 
\multicolumn{4}{l}{{\it non-astrometric parametrizer:}} \\
nSED, (RVS)	& $\Rightarrow$	& \teff, \logg, \fe, & \\ 
	& 	&  \extinct, BC, \aabun ?	& atmospheric model 	\\ 
\\
\multicolumn{4}{l}{{\it additional use of astrometry gives:}} \\
SED, BC, \parallax, \extinct\	& $\Rightarrow$	& L	& $2.5\log L - f(\rm{SED,BC})$\\
&&& \hspace*{1em}$= A - 5\log{\pi}$ \\ 
L, \teff\	& $\Rightarrow$	& R	& $L = 4 \pi R^2 \sigma T_{\rm eff}^4$ \\
\logg, R	& $\Rightarrow$	& M	& $g=GM/R^2$	\\
\\
SED, RVS, \vv(t), \rv(t)	& $\Rightarrow$	& detect unresolved binaries	& orbital model \\
SED(t), RVS(t)	 		& $\Rightarrow$	& detect variables		& variability model \\
\hline
\end{tabular}
\caption[]{Stellar parameters derivable from the GAIA data.
SED=spectral energy distribution (15 photometric measures in medium
and broad band filters); nSED=normalized SED (absolute flux
information removed); RVS=radial velocity spectrum; BC=bolometric
correction; \parallax=parallax; \extinct=interstellar extinction
function; \vv(t) \& \rv(t)=point source velocity and position as a
function of time (from c.\ 70 observations over four years).}
\label{parameters}
\end{table}

Most work on stellar parametrization (and MK classification) has
relied on high resolution spectra from which \teff, \logg\ and \fe\
have been determined (see \cite{bailer-jones_2001} for a review). GAIA
is rather different in that it observes at lower spectral resolution
but measures absolute fluxes as well as parallaxes.
Table~\ref{parameters} shows how stellar parameters can in principle
be derived from these data.  The distance measurement precision for
V=15 is 0.5\% at 500pc, 1\% at 1kpc and 5\% at 5kpc. At V=18 these are
about 2\%, 4\% and 20\% respectively. (These improve by a factor of
two or more for late-type or very reddened stars.) For V=15 the SNR
per filter (at end of mission) is better than 200, decreasing to
50--150 at V=18 and 10--50 at V=20 for most filters and spectral types
\cite{esa_2000}. Thus some 20 million stars will have their distance
determined to better than 1\% and have high precision SEDs. If \teff\
can be established to 1\% then the radii of many of these stars is
determinable to within 2\%. Even in those more frequent cases where
distance determination is too poor to permit a precise radius, it will
provide an approximate intrinsic luminosity which, when combined with
the effective temperature, can be used to aid surface gravity
determination. If \logg\ can be measured to 0.2~dex, then provided R
can be established to within 10\%, a mass determination to within 50\%
is possible without calibration from binary systems. Although poor for
an individual star, it becomes statistically meaningful for a large
sample of similar stars, one of GAIA's strong points. Better
individual masses will be possible from calibration using the 65\,000
visual binaries observed by GAIA for which masses should be obtained
to within 10\% (or 17\,000 to withint 1\%) \cite{esa_2000}. Ages
(possibly with large uncertainties) can be quantified from
evolutionary models.


Being a deep, all-sky survey, GAIA will have to account for
interstellar extinction and its variation across the Galaxy. Moreover,
in order to determine the stellar luminosity and hence radius from the
SED and the parallax, a precise knowledge of the interstellar
extinction is necessary: to determine R (radius) to 2\%, the
extinction must be measured to within 0.03 mags.

When trying to determine several astrometric parameters from a
dataset there exists the problem of parameter degeneracy, i.e.\ two
different astrophysical parameters manifesting themselves in the same
way in the SED in certain parts of the astrophysical parameter
space. An example is \teff\ and extinction in late-type stars, where
lowering \teff\ has a similar effect on the SED (at low resolution) as
increasing the extinction. (The radial velocity spectrum will
help for the brighter stars as this reddening-free information provides an independent
measure of the stellar parameters.)  A more subtle example of
degeneracy is the effect of higher temperatures on metallicity
determination: in hot stars the metals are ionized leaving only very
weak metal lines, making it difficult to determine metallicity in O
and B stars at low resolution. There is then the danger that a
parametrization algorithm could confuse metallicity and temperature
characteristics.  It is therefore essential that these parameters are
determined simultaneously. Clearly, for degenerate cases, a
parametrization algorithm is required which can give a range of
possible parameters, and not just a single set.

Most stellar systems consist of more than one component. Undetected
binaries bias the parameter determinations when the brightness ratio
is small (e.g.\ a higher luminosity for a given \teff\ leads to an
erroneous \fe\ determination).  As GAIA will observe each object
approximately 100 times over a period of four years, about 35\% of all
unresolved binaries out to 1kpc with V$\leq$20 should be
astrometrically detectable \cite{esa_2000}.
The radial velocity can similarly be used.  Upper limits can also be
placed on the mass ratio/separation of any companions. In many cases,
however, the dynamical information will not reveal any useful
information about a companion (e.g.\ for distant, widely separated
binaries, or non-physical companions). In these cases, parametrization
techniques are required which can identify binary stars from their
composite SEDs and ideally parametrize both components.

\section{Practical requirements of a parametrization system}
\label{pracreq}

GAIA consists of three separate telescopes with their viewing angles
lying in the same plane, but separated by 127\deg, 127\deg\ and 106\deg. 
The satellite
rotates once every three hours about an axis perpendicular to this
plane; a slow precession of this axis permits it to view the whole
sky.  The two identical astrometric telescopes observe all objects in
white light (the G band in the AF -- astrometric field) and four broad band
filters (in the BBP instrument). The spectroscopic telescope acquires
the radial velocity spectrum (RVS) plus photometry in 15 broad and
medium band filters (MBP).
All instruments use CCDs in time-delayed integration (TDI) mode
synchronized to the satellite's rotation rate.

GAIA does not have an input catalogue: the astrometric and
spectroscopic instruments have independent star mappers to locate
objects crossing onto the focal plane. All point sources brighter than
some threshold will be observed: not only stars, but also quasars,
galaxies and asteroids. The stellar
parametrization system must therefore be able to identify which objects are
stars. Furthermore, GAIA will inevitably observe rare and previously
unrecognised objects, such as stars in a very brief phase of stellar
evolution.  The efficient identification of such unknown objects is a
task of upmost importance.

Missing data is inevitable in any large survey. For example, the BBP
has a higher spatial resolution than the MBP, so objects which are
resolved in the former may be merged in the latter.  The BBP and
MBP will have a similar magnitude limit of V$\simeq$20 -- although due
to the precession, not all of the same objects are observed in each
instrument in a single satellite rotation -- whereas the RVS will record
spectra only down to about V$\simeq$17. These spectra will blend in
crowded fields, and parts of the spectrum will be lost at the edge of
the field.  Almost all work to date on automated classification has
used `cleaned' data sets in which such problem cases are removed.
However, as GAIA is a complete survey of the sky down to very well
defined magnitude limits, such cleaning would bias statistical
analyses. Worse, it would miss whole classes of objects in the case of
`censored' data, i.e.\ limits on unobserved data, typically an upper
limit on a flux non-detection.  Thus the parametrization
algorithms must be robust to missing data and be able to recognise the
difference between lost and censored data.

The uncertainties in the extracted data -- the SED, RVS and parallaxes
-- can be estimated with some reliability from photon statistics and
detector noise models. It is clearly advantageous if this information
can be relayed to the parametrization algorithm as a measure of
reliability. For fainter and more
distant stars, for example, the parallax gives only an approximate distance,
but it is inefficient to assign an arbitrary magnitude/distance below
which the parallax is no longer used: it will always provide some
constraint, even if weak. A parametrizer which can take advantage of
the error estimates on input data can use these to provide appropriate
uncertainties on the derived parameters. Furthermore, large
uncertainties in the parameters may indicate problems or potentially
interesting objects.

It is often overlooked that prior knowledge of stellar parameters is
sometimes available, and its inclusion can be very beneficial. For
example, the interstellar extinction in parts of the sky can be
estimated from existing extinction maps and data, such as H{\small I}
maps, COBE and 2MASS.  If we independently know, for example, that the
extinction is low, this makes \teff\ determination more reliable in
late-type stars.

\section{Parametrization models}
\label{models}

Further development and testing of suitable parametrization algorithms
will be necessary to meet the above requirements: I will now briefly
comment on the applicability to GAIA of the few automated
techniques which have been used in stellar
parametrization. See \cite{bailer-jones_2001,bailer-jones_etal_2001}
and references therein for more details.

A commonly-used method of classification is the minimum distance
method (MDM), where a class is assigned based on the best matching
template spectrum according to the shortest distance in an
N-dimensional data space (N is the number of spectral elements).
Ideally an interpolation is done through several nearby templates to
allow a continuous parametrization.  $\chi^2$ minimization and
cross-correlation are special cases of MDM and the k-nearest neighbour
(k-nn) method is closely related.  Neural networks can be used
for parametrization by giving a functional mapping between the spectra
at its inputs and the parameters at its outputs.  The optimal mapping
is found by training the network (i.e.\ setting its weights) on a set
of pre-classified spectra (templates). There is a close relationship
between all of these methods: Neural networks perform a {\it global}
interpolation of the training data to come up with a single function
of the parameter(s) in terms of the spectral data. MDM and k-nn, on
the other hand, do a {\it local} interpolation of the training data
every time a new object is to be parametrized (although often authors
dispense with the interpolation and simply assign the parameters of
the nearest template). MDM and k-nn differ in that MDM does
its interpolation in the parameter space, whereas k-nn -- like neural
networks -- does an interpolation in the data space.

Although rarely done in the literature, all of these methods can
accommodate errors in the input data and assign uncertainties to the
derived stellar parameters. MDM and k-nn should be set up to determine
parameters simultaneously to avoid unnecessary problems with parameter
degeneracy. It is unclear whether censored data can be handled by
these methods (although missing data should present no problem) and
whether prior information can be used effectively.  Ideally, a
suitable weighting should be used to establish the optimal distance
measure, e.g.\ from a numerical optimization using the training data.
With neural networks, multiple outputs ensure simultaneity of
parameter determinations from the fact that some of the network
weights are shared by the different outputs. Priors can be
incorporated, although not always in the most useful way. Being a
global interpolator, the frequency distribution in the training data
affects the solution for the whole parameter space, which can be seen
as a drawback.

There are various probabilistic approaches to classification in the
literature. These have the advantage that errors and inputs
can be dealt with in a consistent manner, and that degenerate
(multimodal) solutions can be handled naturally.  A Bayesian framework
is appropriate because it enables us to specify prior information and
its uncertainty explicitly 
(e.g.\ the extinction is $2.0 \pm 0.5$ mags) and treat parametrization
as a learning procedure in which this prior is improved by the data. Most
probabilistic methods used in the literature assign discrete classes
rather than continuous parameters. The latter is obviously more
appropriate for physical parameters which are naturally continuous.
Probabilistic approach to interpolations, such as Gaussian Processes
\cite{bailer-jones_etal_1999} in principle offer the advantages of
both approaches.

\section{A framework solution}
\label{solution}

\begin{figure}[t]
\centerline{
\psfig{figure=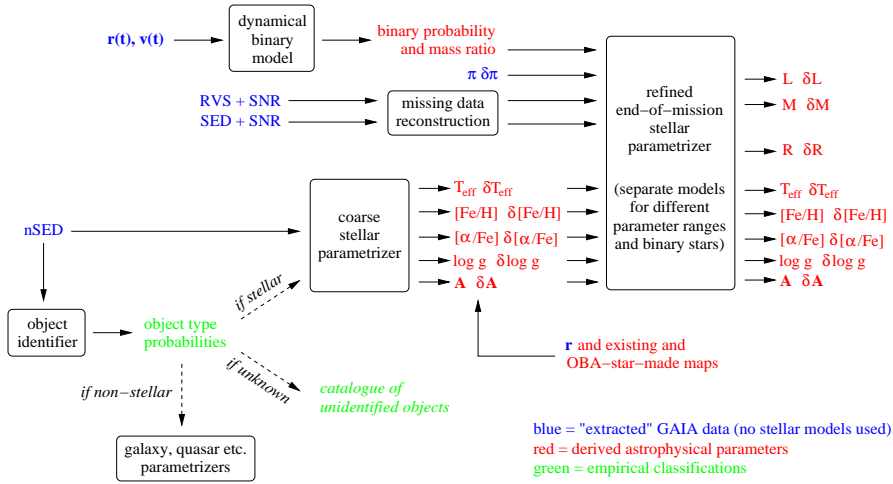,angle=0,width=1.0\textwidth}
}
\caption{A possible approach to the determination of physical stellar
parameters with GAIA (or other astrometric satellite). See
Table~\ref{parameters} for an explanation of the abbreviations.  For
clarity, not all elements of the system are shown. For example,
parallax and proper motion information which only become available at
the end of the mission are useful for identifying extragalactic
objects, and variability is an important means of identifying a number
of types of stars. The whole process would be applied iteratively.}
\label{framework}
\end{figure}

Fig.~\ref{framework} outlines a framework for object parametrization
with GAIA. The identification and parametrization of all types of
stars, galaxies, quasars and asteroids is almost certainly too complex
for a single parametrizer.  In this approach, objects are first
identified using just the colour information in the normalized SED
(nSED). Once the stars have been identified, they are coarsely
parametrized, again using only the colour information.  The idea
behind this is to use a reduced (and usually available) subset of the
total data to make only approximate determinations of the physical
parameters. This coarse parametrizer can therefore be relatively
simple and avoid some of the problems outlined earlier.  These
approximate parameters then serve as the initial estimates to guide a
more `refined' and sophisticated parametrizer, which can take account
of additional information more reliably.

The refined classifier has a number of additional features. For
example, missing spectral data are reconstructed statistically, using
the correlations present in the complete data cases.  This does not,
of course, add information in individual cases, but does provide a way
for the parametrizer to take advantage of redundancy in the data. The
refined parametrizer should also use estimates of the mass ratio of
possible unresolved binaries to look for evidence of two components in
the SED.

The refined parametrizer can incorporate other prior information,
e.g.\ external extinction data from other surveys.  Furthermore, where
the interstellar extinction varies slowly through space (away from
star forming regions and interstellar clouds), it can be measured much
more reliably from B, A and F stars than from later type stars. Thus
the refined parametrizer could first be applied to early-type stars to
map the extinction in three dimensions. More generally, an iterative
application of the whole process in Fig.~\ref{framework} allows the
missing data reconstruction algorithm to learn from the complete data
cases and thus make more reliable parametrizations in cases with
missing data.

The determination of physical parameters relies ultimately on stellar
models and synthetic spectra, so these could (at least initially) form
the training data of a parametrizer. Given the parameters, the stellar
models provide a unique spectrum. The goal of parametrization is to do
the inverse (and not necessarily unique) problem of determining the
parameters from a noisy part of this spectrum.  Combining the synthetic
spectra with empirical and theoretical mass-radius-\teff\
relationships and simulating observations of them at different
distances and extinctions, we can assemble up a library of SEDs as a
function of known \teff, \logg, \fe, \aabun, \parallax\ and
\extinct. The mass, radius and luminosity are therefore also known,
albeit within the scatter in the mass--radius--\teff\
relationships. This reliance on stellar models means that existing
models need to be improved to accurately reflect the full range of
stars which GAIA will encounter.  Particular aspects are \aabun\
variations, CNO anomalies, NLTE effects and dust formation. One of the
goals of GAIA is of course to improve such stellar models, and these
can be introduced into the iterative parametrization process via the
training data to achieve a self-consistent solution.  Stellar
parameters can only be assigned to stars for which we have some
notion: it makes no sense to attempt to assign parameters to unknown
types of stars for which the details of SED formation are
unknown. While provision must certainly be made for the identification
and empirical classification of new types of objects, e.g.\ using
unsupervised methods, their physical characterisation must rely on
detailed spectroscopic follow-up.

Ideally, the parametrizers would be trained on real and not simulated
data.  At the very least, synthetic spectra would require broad
observational corrections, e.g.\
\cite{lejeune_etal_1997}. Alternatively, a representative set of stars
across the whole parameter space could be observed at sufficiently
high spectral resolution and their parameters determined, either with
direct physical methods or by the method of Bailer-Jones (1997). These
observed spectra could then be convolved with the GAIA instrumental
model and used to train the parametrizers. Better still, GAIA
observations of these same objects could be taken from the GAIA
database and used as the training data.

While I advocate using as much of the GAIA data as possible for
parametrization, the 6D phase space co-ordinates (position and
velocity) obtained by GAIA should initially be excluded: stars should
first be parametrized according to their {\em intrinsic} properties.
A correlation of these with their Galactic phase space location is of
course one of the major goals of GAIA. But we should initially keep
stellar structure and Galactic structure separate: while the
properties of a star may be correlated and physically {\em related} to
its phase space location, they are not directly {\em caused} by it, so
an independent parametrization is desirable.

\section{Summary}
\label{summary}

I have described some of the problems faced in determining physical
parameters of the one billion stars which will be observed with GAIA.
The scientific requirements for this can be summarized as follows:
determination of \teff, \logg, \fe, \extinct\ (interstellar
extinction), and, where possible, \aabun, CNO abundance anomalies,
\vrot, \vmicro, radius, mass and age, plus error estimates on all of these;
identification of degenerate cases; identification and
parametrization of unresolved binaries, again with error estimates;
explicit identification of strange objects (maybe new types of
objects).  The practical requirements are: initial identification of
stars (ideally with a probability assigned); use of all available data
(RVS, parallax, dynamical and variability data in addition to the
SED); advantage made of error estimates of input data; be robust to
missing and censored data; use prior information where available. A
framework solution was suggested in which parametrization proceeds in
three stages: object identification; coarse parametrization; refined
parametrization. It uses the parallax and local astrometry in addition
to the spectral energy distribution. The whole procedure can be
iteratively applied to the end-of-mission data to improve the
parameter determinations and handling of missing data and to permit
the incorporation of improved stellar models.

\end{article}

\begin{thebibliography}{}

\bibitem[\protect\citeauthoryear{Bailer-Jones}{2001}]{bailer-jones_2001}
Bailer-Jones C.A.L., 
\newblock {Automated stellar classification for large
surveys: a review of methods and results},
\newblock in {\it Automated Data Analysis in Astronomy}, 
R.\ Gupta, H.P.\ Singh, C.A.L.\ Bailer-Jones (eds.), 
Narosa Publishing House, New Delhi, India, pp.~83--98, 2001

\bibitem[\protect\citeauthoryear{Bailer-Jones et al.}{1997}]{bailer-jones_etal_1997}
Bailer-Jones C.A.L., Irwin M., Gilmore G., von Hippel T., 
\newblock {Physical parameterization of stellar spectra: The neural network
approach},
\newblock MNRAS, 292, 157--166, 1997

\bibitem[\protect\citeauthoryear{Bailer-Jones et al.}{1999}]{bailer-jones_etal_1999}
Bailer-Jones C.A.L., Bhadeshia H.K.D.H., MacKay D.J.C.,
\newblock {Gaussian process modelling of austenite formation in steel},
\newblock Materials Science and Technology, 15, 287--294, 1999

\bibitem[\protect\citeauthoryear{Bailer-Jones et al.}{2001}]{bailer-jones_etal_2001}
Bailer-Jones C.A.L., Gupta R., Singh H.P., 
\newblock {An introduction to artificial neural networks}, 
\newblock in {\it Automated Data Analysis in Astronomy}, 
R.\ Gupta, H.P.\ Singh, C.A.L.\ Bailer-Jones (eds.), 
Narosa Publishing House, New Delhi, India, pp.~51--68, 2001

\bibitem[\protect\citeauthoryear{ESA}{2000}]{esa_2000}
ESA,
\newblock {GAIA: Composition, formation and evolution of the Galaxy},
\newblock ESA-SCI(2000)4, 2000

\bibitem[\protect\citeauthoryear{Lejeune et al.}{1997}]{lejeune_etal_1997}
Lejeune T., Cuisinier F., Buser R., 
\newblock  {A standard stellar library for evolutionary synthesis.\ 
I.\ Calibration of theoretical spectra,}
\newblock A\&AS, 125, 229--246, 1997

\bibitem[\protect\citeauthoryear{Perryman et al.}{2001}]{perryman_etal_2001}
Perryman M.A.C. et al., 
\newblock {GAIA: Composition, formation and evolution of the Galaxy}
\newblock A\&A, 369, 339--363, 2001

\end{thebibliography}
\end{document}